\documentclass[%
 reprint,
superscriptaddress,
 amsmath,amssymb,
 aps,prl
]{revtex4-2}

\usepackage{graphicx}
\usepackage{dcolumn}
\usepackage{bm}
\usepackage{bbm}
\usepackage{amsmath,amsthm,amssymb,bm}
\usepackage{physics}
\usepackage{old-arrows}
\usepackage{leftindex}
\usepackage{tikz}
\usetikzlibrary{shapes}
\usepackage[normalem]{ulem} 
\usepackage{xcolor}
\definecolor{customblue}{HTML}{D6E8F5}

\makeatletter
\newdimen\@myBoxHeight%
\newdimen\@myBoxDepth%
\newdimen\@myBoxWidth%
\newdimen\@myBoxSize%
\newcommand{\SquareBox}[2][]{%
    \settoheight{\@myBoxHeight}{#2}
    \settodepth{\@myBoxDepth}{#2}
    \settowidth{\@myBoxWidth}{#2}
    \pgfmathsetlength{\@myBoxSize}{max(\@myBoxWidth,(\@myBoxHeight+\@myBoxDepth))}%
    \tikz \node [shape=rectangle, shape aspect=1,draw=red,inner sep=2\pgflinewidth, minimum size=\@myBoxSize,#1] {#2};%
}%
\makeatother

\newcommand{\ah}{ \hat{a} }
\newcommand{\ad}{ \hat{a}^\dag}
\newcommand{\Var}{ \textrm{Var}}

\begin{document}

\preprint{APS/123-QED}

\title{Discrete-variable assisted error correction of continuous-variable quantum information}

\author{Negin Razian}
 \altaffiliation{negin\_razian@sfu.ca}
 \affiliation{Department of Physics, Simon Fraser University, Burnaby, British Columbia, Canada V5A 1S6}

\author{En-Jui Chang}
\affiliation{Department of Physics, Simon Fraser University, Burnaby, British Columbia, Canada V5A 1S6}

\author{Hoi-Kwan Lau}
 \email{hklau.physics@gmail.com}
 \affiliation{Department of Physics, Simon Fraser University, Burnaby, British Columbia, Canada V5A 1S6}

\date{\today}

\begin{abstract}
Robust continuous-variable (CV) quantum information processing requires correcting realistic errors in bosonic systems, but all existing schemes rely on auxiliary Gottesman-Kitaev-Preskill (GKP) states which the preparation and operation are demanding in many platforms. In this work, we propose a novel CV quantum error correction (QEC) scheme that utilizes a broadly accessible resource: discrete-variable (DV) ancilla. Our scheme extracts information about CV displacement to the DV ancilla, measuring that allows counteracting the unwanted displacement error. We show that a simple single-qubit ancilla can already suppress CV infidelity by more than $20\%$. By concatenating with DV QEC codes, our scheme is robust against the physical errors in hybrid CV-DV systems, and yields a new class of oscillator-in-oscillator code that does not involve GKP states. Our work facilitates the implementation of CV QEC on realistic platforms.
\end{abstract}
\maketitle

\textit{Introduction}---Harnessing quantum effects can enhance the performance of many technological applications such as computation, communication, and sensing \cite{NielsenChuang}. Among different implementations of quantum technologies, the bosonic platforms, e.g. photons and mechanical oscillators, are particularly promising \cite{Andersen:2015dp}. Some bosonic systems offer superior physical properties, for example, high mobility \cite{Yin:2017ij}, long coherence time \cite{coherence,10.1126/science.abc7312} and scalable entanglement generation \cite{Yokoyama:2013ux,2016APLP....1f0801Y,2019arXiv190608709L,10.1126/science.aay2645, 10.1103/physrevlett.130.120601, 10.1103/physrevlett.134.183602}. Furthermore, each bosonic degree of freedom (mode) contains essentially infinite quantum states, the huge information capacity allows quantum information to be encoded as either a superposition of discrete-variable (DV) basis states or a continuous-variable (CV) wavefunction \cite{cvadvantages_Braunstein,cvadvantages_Weedbrook}. In particular, the CV approach has found applications in quantum simulation \cite{cv_quantum_computation, simulation1, Marshall:2015fx}, demonstrating quantum computational advantages \cite{bosonsampling2}, machine learning \cite{Lau:2017ky,10.1103/physrevresearch.1.033063}, and many more \cite{10.1103/4rf7-9tfx}.

Nevertheless, most quantum technologies today have been developed with the DV approach. One reason is that quantum error correction (QEC), which is essential to protect quantum information against inevitable environmental influence, is much less developed for CV than the DV counterpart. Early proposals of CV QEC attempted to directly generalize the DV schemes \cite{cvqec1,cvqec2,cvqec3,cvqec4}, but they cannot offer protection against the errors in realistic bosonic platforms.
A breakthrough scheme was introduced by Noh \textit{et.~al.} \cite{Jiang} that utilizes auxiliary modes in Gottesman-Kitaev-Preskill (GKP) states \cite{gkp}. By correlating the noises in the system and auxiliary modes and using the high sensitivity of the GKP states to disturbance in both quadratures \cite{2017PhRvA..95a2305D}, this ``oscillator-in-oscillator code" can suppress the naturally occurring random displacement noise. Since then, the GKP-based scheme has been extended to correct a wide range of realistic errors \cite{ostos1,ostos4,10.22331/q-2023-08-16-1082}.

Despite the promises, the implementation of GKP-based CV QEC faces several technical difficulties. First, although preliminary GKP states have been realized \cite{10.1038/s41586-019-0960-6,10.1038/s41586-020-2603-3,10.1126/science.adk7560}, the deterministic generation of high-quality GKP states remains challenging on most bosonic platforms. Furthermore, in these schemes extracting the error syndromes requires CV quadrature measurement. Although this can be routinely implemented in optical platforms by homodyne detection, it is difficult to realize this on hybrid DV-CV platforms, whose mode properties are usually measured through binary measurements of the companion qubit. Therefore, a CV QEC scheme without GKP state is practically desired.

In this work, we introduce a novel CV QEC scheme that incorporates a highly accessible resource: auxiliary DV systems. 
Our scheme utilizes the DV ancilla to extract information about the CV noise, so its influence can be subsequently counteracted. We show that an ancilla as simple as a single qubit can already suppress the infidelity of a noisy CV state by over 20\%. Furthermore, by concatenating DV QEC to protect the ancilla, our scheme is effective against errors in a broad range of hybrid or purely bosonic platforms.
In particular, by using bosonic codes, our scheme introduces a new class of oscillator-in-oscillator code that does not require GKP states. Our work provides new flexibilities to tailor CV QEC according to the properties of bosonic platforms.

\textit{CV quantum error}---In this work we focus on the random displacement error. This can be commonly found in bosonic platforms, for example, due to electric field fluctuations in ion trap platforms \cite{2015RvMP...87.1419B} or as an overall effect of photon loss after being compensated by amplification \cite{10.1109/tit.2018.2873764}. The error will transform an initial CV state $\hat{\rho}_{cv}$ as
\begin{equation}\label{eq:noise_model}
    \hat\rho_{cv} \longrightarrow \mathcal{E}(\hat\rho_{cv})\equiv \int P(\beta) \hat{D}(\beta) \hat\rho_{cv} \hat{D}^\dagger(\beta) d^2\beta~,
\end{equation}
where $\hat{D}(\beta) = \exp(\beta \hat{a}^\dagger - \beta^* \hat{a})$ is the displacement operator; $\beta = \beta_q + i \beta_p$, which $\beta_q$ and $\beta_p$ are respectively the unknown displacement along $q$ and $p$ quadratures; $P(\beta)$ is the distribution of the unknown displacement; $d^2\beta$ is the shorthand for $d\beta_q d\beta_p$; the mode operators obey $[\ah,\ad]=1$. For a physical process that is phase insensitive, the displacement distribution is usually Gaussian and has the same variance in both quadratures, i.e.,
\begin{equation}\label{eq:noise_dist}
    P(\beta) = G(\beta_q,\sigma)G(\beta_p,\sigma)~,
\end{equation}
where $G(x,\sigma)\equiv e^{-x^2/\sigma^2}/\sqrt{\pi}\sigma$. The magnitude of noise is characterized by $\sigma$, which is assumed to be a known parameter, for example, from system characterization.

Intuitively, the error is suppressed if one can reduce the variances of the random displacement. Explicitly, consider a pure initial state $\hat{\rho}_\textrm{cv} = \ket{\psi_\textrm{cv}} \bra{\psi_\textrm{cv}}$. The state fidelity after error is $\mathcal{F}= \int P(\beta) f(\psi_\textrm{cv},\beta)d^2 \beta$, where $f(\psi, \beta) \equiv |\langle \psi | \hat{D}(\beta) | \psi\rangle|^2$ is the state overlap after a displacement $\beta$. If the displacement error is small, the fidelity can be approximated as \cite{supp}
\begin{equation}\label{eq:fidelity_expansion}
   \mathcal{F}\approx 1+\frac{1}{2}\Big( \frac{\partial^2 f}{\partial\beta_q^2} \Big|_{\beta=0} \Var(\beta_q)+\frac{\partial^2 f}{\partial\beta_p^2} \Big|_{\beta=0} \Var(\beta_p) \Big)~,
\end{equation}
for any distribution $P(\beta)$ that is uncorrelated in $q$ and $p$ quadratures and is zero-mean in both quadratures.
Therefore, the infidelity, $1-\mathcal{F}$, is linearly dependent on the variances of $P(\beta)$, which are given by $\Var(\beta_x)\equiv \int P(\beta)\beta^2_x d^2\beta$ for both quadratures $x= \{q,p\}$.  We note that the double derivatives are strictly negative and depend on the state $\ket{\psi_\textrm{cv}}$. For example, both are equal to $-2$ for coherent states and $-6$ for single-boson state $\ket{1}$. For general $\ket{\psi_\textrm{cv}}$, we will quantify the efficacy of CV QEC by the sum of variances, $\Var(\beta_q)+\Var(\beta_p)$ \cite{supp}.

\begin{figure} 
    \centering
    \includegraphics[width=8cm]{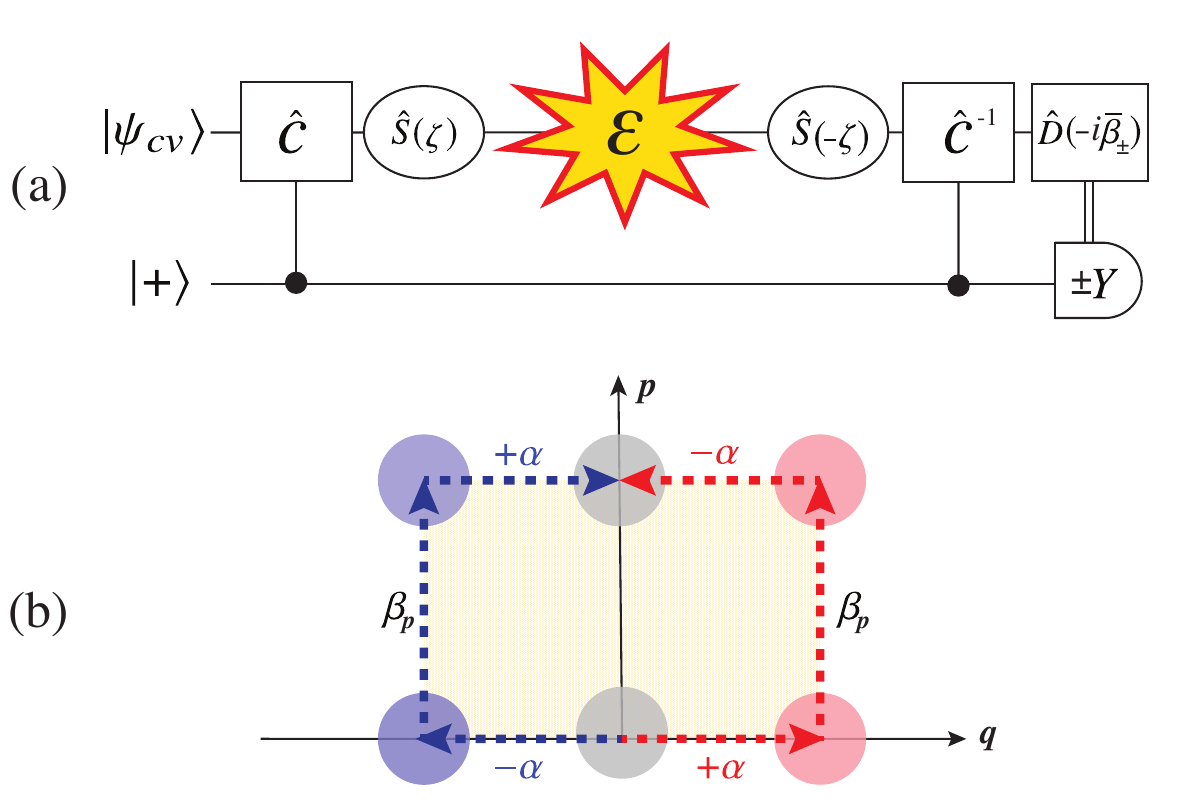}
    \caption{ (a) Circuit of our CV QEC scheme by using a perfect qubit ancilla. 
    (b) Phase-space trajectories of CV state ($\ket{0}$ is shown as an example) during the encoding, $p$-displacement noise, and decoding processes. The trajectory conditional on the qubit state $\ket{g}$ ($\ket{e}$) is shown in blue (red). The area enclosed by the trajectories (shaded yellow) manifests as a geometric phase between the qubit basis states.
    }
    \label{fig:circuit}
\end{figure}

\textit{Suppressing single-quadrature noise}---Our CV QEC aims to use a DV ancilla to obtain information about the unknown displacement $\beta$, so to subsequently counteract the displacement and suppress the displacement variance. To illustrate the principle of our scheme, we first assume that the DV ancilla is a perfect qubit and only one quadrature, say $p$, suffers from noise, i.e. $P(\beta)=G(\beta_p,\sigma)$. 
Our scheme is illustrated in Fig.~\ref{fig:circuit}(a) (neglecting the squeezing operations at this moment). It begins by preparing the qubit in an equal superposition of basis states, $|+\rangle = (|g\rangle + |e\rangle)/\sqrt{2}$. The encoding process is to apply a conditional-displacement $\hat{\mathcal{C}}$ to the joint qubit-mode system
\begin{equation}\label{eq:conditional displacement}
    \hat{\mathcal{C}} = \ket{g}\bra{g} \otimes\hat{D}(-\alpha) + \ket{e}\bra{e} \otimes\hat{D}(\alpha)~,
\end{equation}
where $\alpha$ is real, so the conditional displacement is along $q$ quadrature. The bosonic mode then undergoes random displacement error (\ref{eq:noise_model}). Afterwards, the system is decoded by inverting the conditional-displacement, $\hat{\mathcal{C}}^{-1}$.  

As illustrated in Fig.~\ref{fig:circuit}(b), after the above process the CV state will be displaced by an unknown $\beta_p$, just as there is no en/decoding. However, conditional on the qubit state, the CV state follows different trajectories. A geometric phase proportional to the phase-space area enclosed by the two trajectories will be induced onto the qubit \cite{10.1038/nphoton.2013.172,10.22331/q-2023-05-31-1024}. The joint state after decoding becomes
\begin{equation}\label{eq:geometric_phase}
    \ket{+} \otimes\ket{\psi_{cv}} \longrightarrow \frac{1}{\sqrt{2}} \left( \ket{g} + e^{i 4\alpha\beta_p} \ket{e} \right)\otimes \hat{D}(\beta_{p}) \ket{\psi_{cv}}~.
\end{equation}
The geometric phase, $4 \alpha \beta_p$, is proportional to the displacement, so information about the displacement can be obtained by measuring the qubit. 

The qubit is then measured in the basis $\{\ket{\pm Y}\equiv(\ket{g}\pm i\ket{e})/\sqrt{2} \}$.  
Upon the measurement outcome $\ket{\pm Y}$, the displacement distribution is modified by a filter function, i.e. $P(\beta)\rightarrow G(\beta_p,\sigma)( 1 \pm \sin(4\alpha\beta_p))$. The modified distribution has mean $\bar{\beta}_\pm \equiv \int \beta_p G(\beta_p,\sigma^2) ( 1 \pm \sin(4\alpha\beta_p)) d\beta= \pm 2\alpha\sigma^2 \exp(-4\alpha^2\sigma^2)$. The correction is thus to displace the CV state back by $\bar{\beta}_\pm$, i.e. applying $\hat{D}(-i \bar{\beta}_{\pm})$.  The corrected displacement distribution becomes
\begin{equation}\label{eq:shifted_dist}
P_\textrm{corr}(\beta) = G(\beta_p+\bar{\beta}_\pm,\sigma)  \Big( 1 \pm \sin\big(4\alpha \big(\beta_p+\bar{\beta}_\pm) \big) \Big).
\end{equation}
We note that the corrected distributions for the two outcomes $\ket{\pm Y}$ are mirror-symmetric about the origin, so their variances are the same.

\begin{figure} 
    \centering
    \includegraphics[width=8.5cm]{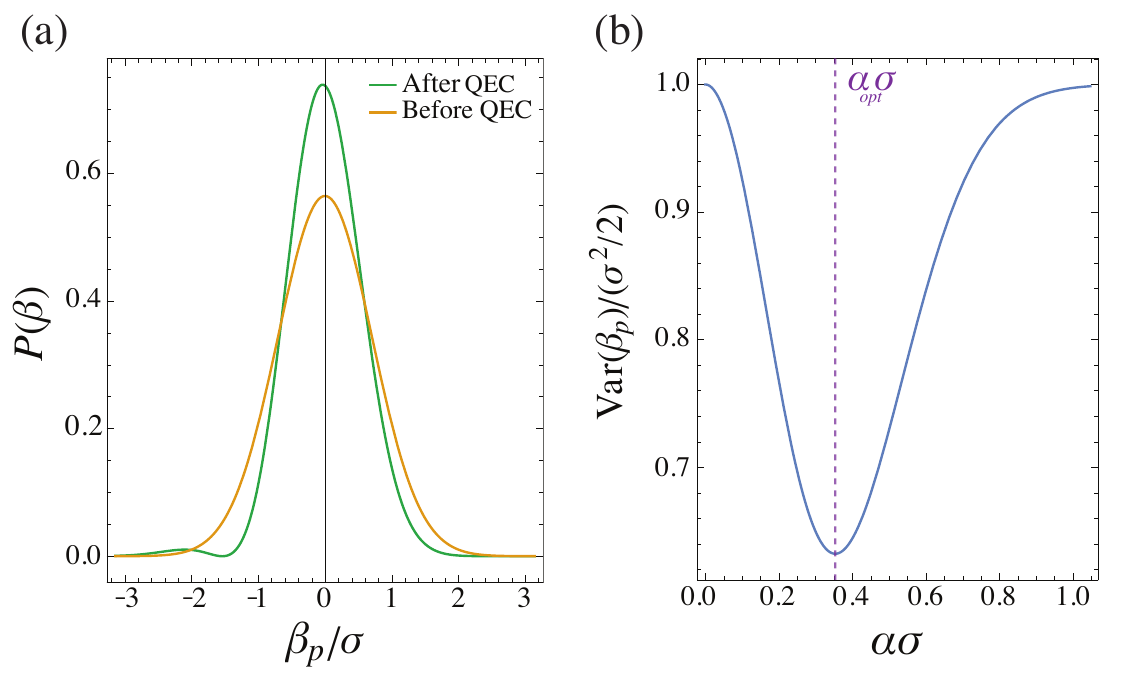}
    \caption{\small (a) Distribution of $p$-quadrature random displacement before (yellow) and after (green) CV QEC with a single-qubit ancilla. Only the corrected distribution of $\ket{+Y}$ outcome is shown; the $\ket{-Y}$ outcome distribution is a mirror symmetry and has the same variance.
    (b) Corrected displacement variance against different choice of conditional-displacement strength $\alpha$.
}
    \label{fig:Distribution}
\end{figure}

The performance of CV QEC depends on the conditional-displacement strength $\alpha$. Small $\alpha$ would produce a small geometric phase and thus reduce the displacement information obtained from qubit measurement. On the other hand, the same geometric phase is produced by the displacements $\beta_p + n\pi/2\alpha$ for any integer $n$. If $\alpha$ is too large, several of these displacements would be probable within the original Gaussian distribution $G(\beta_p,\sigma)$, so the qubit measurement result cannot provide an accurate estimation of the displacement. As shown in Fig.~\ref{fig:Distribution}(b), we find the optimum value that minimizes the displacement variance is $\alpha_{opt} = \frac{1}{2 \sqrt{2}\sigma}$. The corrected distribution, shown in Fig.~\ref{fig:Distribution}(a), has a variance $\Var(\beta_p)=(1-1/e)\sigma^2/2$, which is 36.8\% smaller than that of the original distribution, $\sigma^2/2$. 

\textit{Suppressing both quadrature noises}---Now we return to the realistic noise (\ref{eq:noise_dist}) that randomly displaces in both $q$ and $p$ quadratures. The single-qubit scheme described above can correct only $p$- but not $q$-quadrature noise. It is because the conditional-displacement (\ref{eq:conditional displacement}) commutes with $q$ displacement, no geometric phase is induced. A natural modification to our scheme is to incorporate two ancilla qubits, one for correcting $p$ noise and the other for $q$ noise; the latter can be implemented with a conditional-displacement along $p$ quadrature, i.e. Eq.~(\ref{eq:conditional displacement}) with an imaginary $\alpha$. Then the displacement variance in each quadrature, and hence the total variance $\Var(\beta_q)+\Var(\beta_p)$, will be suppressed by 36.8\% \cite{supp}.

Alternatively, with a simple modification, a single-qubit ancilla can offer protection against \textit{both} quadrature noises. The idea is to squeeze the mode before it undergoes error. If the squeezing amplifies the CV state along $q$ quadrature, the state becomes less susceptible to $q$ noise. As a tradeoff, the CV state is deamplified in $p$ quadrature and experiences more $p$ noise. However, $p$ noise can be suppressed by our single-qubit CV QEC, so both quadrature noises are eventually reduced. This idea can be executed by applying squeezing, $\hat{S}(\zeta) = \exp( \zeta \hat{a}^2 - \zeta \hat{a}^{\dagger 2} )$, and anti-squeezing, $\hat{S}(-\zeta)$, to the mode before and after the error, see Fig.~\ref{fig:circuit}(a). 

We analytically find that the total displacement variance is minimum when $\zeta = \frac{1}{8} \ln(1 - e^{-1})$, which corresponds to $0.996$~dB squeezing. This amount of squeezing produces the same corrected variance in both quadratures, i.e. $\text{Var}(\beta_q)=\text{Var}(\beta_p)$.  The total variance is reduced by a factor of $1-\sqrt{1-1/e}\approx 20.5\%$ \cite{supp}.

\textit{Qudit ancilla---}Since the ancilla qubit measurement gives only one bit of information, it cannot provide the full description of the displacement $\beta$, which is a continuous number. A natural resolution is to increase the dimension of the DV ancilla. We generalize our scheme to incorporate a qudit ancilla with dimension $d$ and basis $\{\ket{g_k} \}$. The qudit is initialized as an equal superposition of all basis states, $\sum_{k=1}^d \ket{g_k}/\sqrt{d}$. Encoding process can be implemented by a generalized conditional-displacement, $\hat{\mathcal{C}}_d = \sum_{k=1}^d |g_k\rangle\langle g_k| \otimes \hat{D}(k \alpha)$.  Here we choose $\alpha$ to be real and focus on suppressing $p$ noise; the scheme can be extended to suppress also $q$ noise by using squeezing, as discussed earlier. After noise and decoding by $\hat{\mathcal{C}}_d^{-1}$, each basis state acquires a geometric phase, i.e. $\sum_{k=1}^d \ket{g_k}/\sqrt{d}\rightarrow \sum_{k=1}^d e^{-i2k \alpha \beta_p}\ket{g_k}/\sqrt{d}$. The qudit is then measured in the Fourier basis, i.e. $\{ \ket{+_l}\equiv \sum_{k=1}^d \exp(-i 2 \pi k l/d) \ket{g_k}/\sqrt{d} \}$. 

\begin{figure} 
    \centering
    \includegraphics[width=8.5cm]{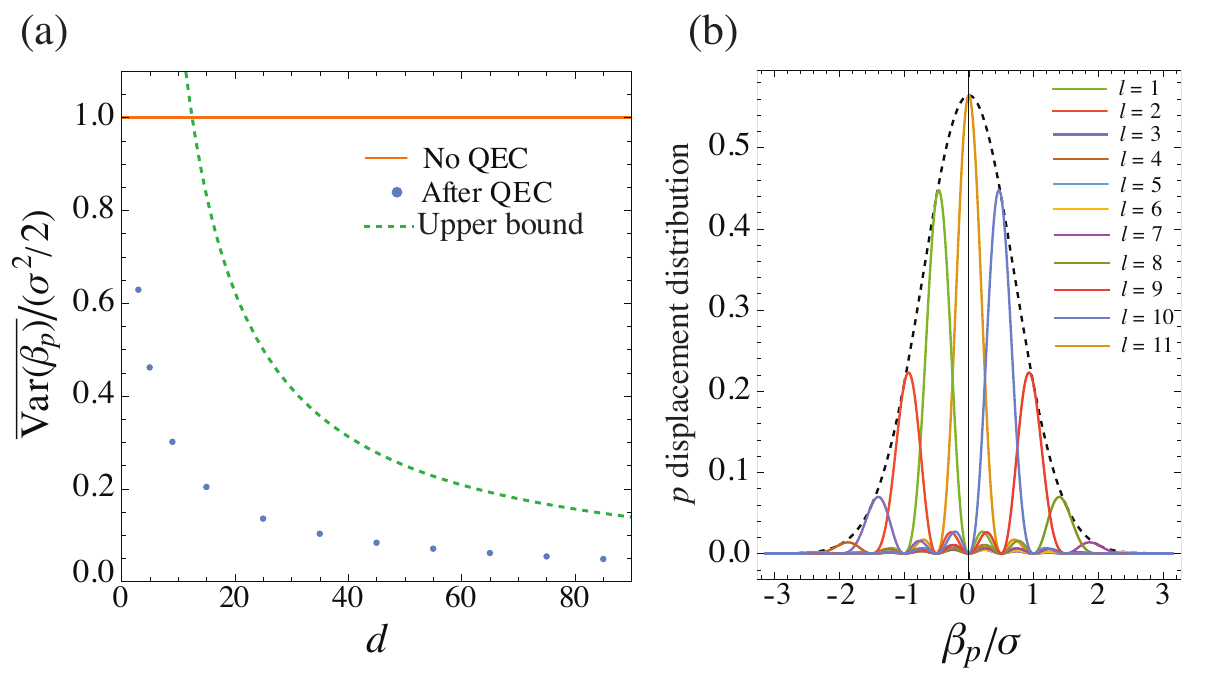}
    \caption{ (a) Average corrected $p$-displacement variance (blue dots) in Eq.~(\ref{eq:var_beta_d}) against qudit ancilla dimension. The upper bound (green dashed) is given by Eq.~(\ref{eq:qudit_bound}) with $s=5$.
    (b) Unnormalized displacement distribution (\ref{eq:qudit_p_dist}), $\mathcal{N}_l P_l(\beta_p)$, after obtaining outcome $l$. Here $d=11$. Dashed line shows the original Gaussian distribution.
    }
    \label{fig:qudit}
\end{figure}

Upon obtaining an outcome $\ket{+_l}$, the distribution of $p$ displacement is modified as (illustrated in Fig.~\ref{fig:qudit}(b)) \cite{supp}
\begin{equation}\label{eq:qudit_p_dist}
G(\beta_p,\sigma) \rightarrow P_l(\beta_p)\equiv\frac{1}{\mathcal{N}_l} G(\beta_p,\sigma) \frac{\sin^2(d \alpha \beta_p + l\pi)}{d^2 \sin^2 (\alpha \beta_p + l\pi/d)}~,
\end{equation}
where $\mathcal{N}_l$ is the probability of obtaining the outcome $\ket{+_l}$. The CV state is then corrected by displacing backward in $p$ quadrature by the mean of the modified distribution, $\bar{\beta}_l \equiv \int \beta_p P_l(\beta_p) d\beta_p$. Including all possible measurement outcomes, the average $p$-displacement variance after correction is given by 
\begin{equation}\label{eq:var_beta_d}
\overline{\Var(\beta_p)} = \sum_{l=1}^d \mathcal{N}_l \int \beta_p^2 P_l(\beta_p+\bar{\beta}_l) d\beta_p~.
\end{equation}
Fig.~\ref{fig:qudit}(a) shows the $\overline{\Var(\beta_p)}$ for different ancilla dimension $d$. The conditional-displacement strength $\alpha$ is optimized by minimizing $\overline{\Var(\beta_p)}$ for each $d$. As expected, the mean variance is progressively suppressed by using an ancilla of higher dimension.

To understand how the variance suppression scales as $d$, we construct an upper bound for (\ref{eq:var_beta_d}),
\begin{equation}\label{eq:qudit_bound}
\overline{\Var(\beta_p)} < \sigma^2 \frac{s^2}{4d}~,
\end{equation}
where $s$ is related to the chosen $\alpha = \pi/s \sigma$ and assumed to be sufficiently large. Therefore, the corrected displacement variance is at least suppressed by $1/d$. Here we only illustrate the intuition of this scaling and defer the details to Supplementary Materials \cite{supp}. We first recognize that the filter functions in (\ref{eq:qudit_p_dist}), $\frac{\sin^2(d \alpha \beta_p + l\pi)}{d^2 \sin^2 (\alpha \beta_p + l\pi/d)}$, have multiple peaks in $\beta_p$ that are regularly separated by $\pi/\alpha= s \sigma$. To learn accurate displacement information from qudit measurement, we require the peak separation is large enough that only one peak is significant in the modified displacement distribution (\ref{eq:qudit_p_dist}), while all other peaks are exponentially suppressed by the tail of the original Gaussian distribution. If the significant peak is close to the origin, the modified distribution (\ref{eq:qudit_p_dist}) will be dominated by the peak. Because the width of the peak $\propto 1/d$, the integral in (\ref{eq:var_beta_d}) scales at $1/d^2$. Otherwise, if the peak is far from the origin, the Gaussian distribution is suppressed by the tail of the filter function, which scales at $1/d^2$.  Therefore, the integral in (\ref{eq:var_beta_d}) also scales at $1/d^2$. Since $\overline{\Var(\beta_p)}$ involves $d$ integrals in (\ref{eq:var_beta_d}), it will scale as $1/d$.

\textit{Imperfect ancilla}---So far, we assumed the DV ancilla is perfect. However, in reality, physical DV systems unavoidably experience decoherence; the DV error would then transfer to the mode through the decoding process and corrupt the CV state. Fortunately, the ancilla is DV, so it can be protected by established DV QEC. 

The choice of DV QEC scheme depends on the type of decoherence and the relative quality of the physical qudits and modes.  Generally, if the physical qudits have a longer coherence time than the modes, such as in trapped ions \cite{Haffner:2008tg}, then the DV ancilla should be a logical qudit encoded by physical qudits. One possibility is to use the multiple available meta-stable states, such as the hyperfine states of an ion, to represent the logical DV states \cite{10.1103/physrevlett.127.010504}. Another possibility is to represent the logical qudit by multiple qubits, such as using several ion qubits in trapped ion bosonic platforms \cite{Kerotrappedion, Shen:2014dm, 10.1038/s41567-023-01952-5}. 

As a demonstration of principle, we consider the dominant physical errors in most DV platforms: dephasing. Any qubit state $\hat{\rho}_{dv}$ would be transformed as 
\begin{equation}\label{eq:dephasing_channel}
\hat\rho_{dv} \rightarrow (1-p_\phi)\hat\rho_{dv} + p_\phi \hat{Z}\hat\rho_{dv}\hat{Z},\end{equation}
where $p_\phi$ is the probability of phase-flip error and $\hat{Z} \equiv |g\rangle\langle g| - |e\rangle\langle e|$. If a dephased qubit is used as an ancilla, the flipped phase cannot be distinguished from the geometric phase and causes wrong displacement estimation.

To suppress dephasing error, we can use a variant of the 3-qubit code \cite{10.1088/0034-4885/76/7/076001},
\begin{equation}\label{eq:three_qubit}
\ket{g}_L=\frac{\ket{+++}+\ket{---}}{\sqrt{2}}~,~\ket{e}_L=\frac{\ket{+++}-\ket{---}}{\sqrt{2}}~.
\end{equation}
If a phase-flip occurs on any one of the physical qubits, $\ket{+}\leftrightarrow \ket{-}$, that qubit can be identified by parity checks and subsequently corrected \cite{10.1088/0034-4885/76/7/076001}. The phase of the logical qubit is flipped only when two or more physical qubits are phase-flipped, the probability of which is given by $p_\phi'=3p_\phi^2-p_\phi^2$ \cite{supp}. The effective dephasing probability is therefore suppressed when $p_\phi$ is sufficiently small.

\begin{figure}
    \centering
    \includegraphics[width=8cm]{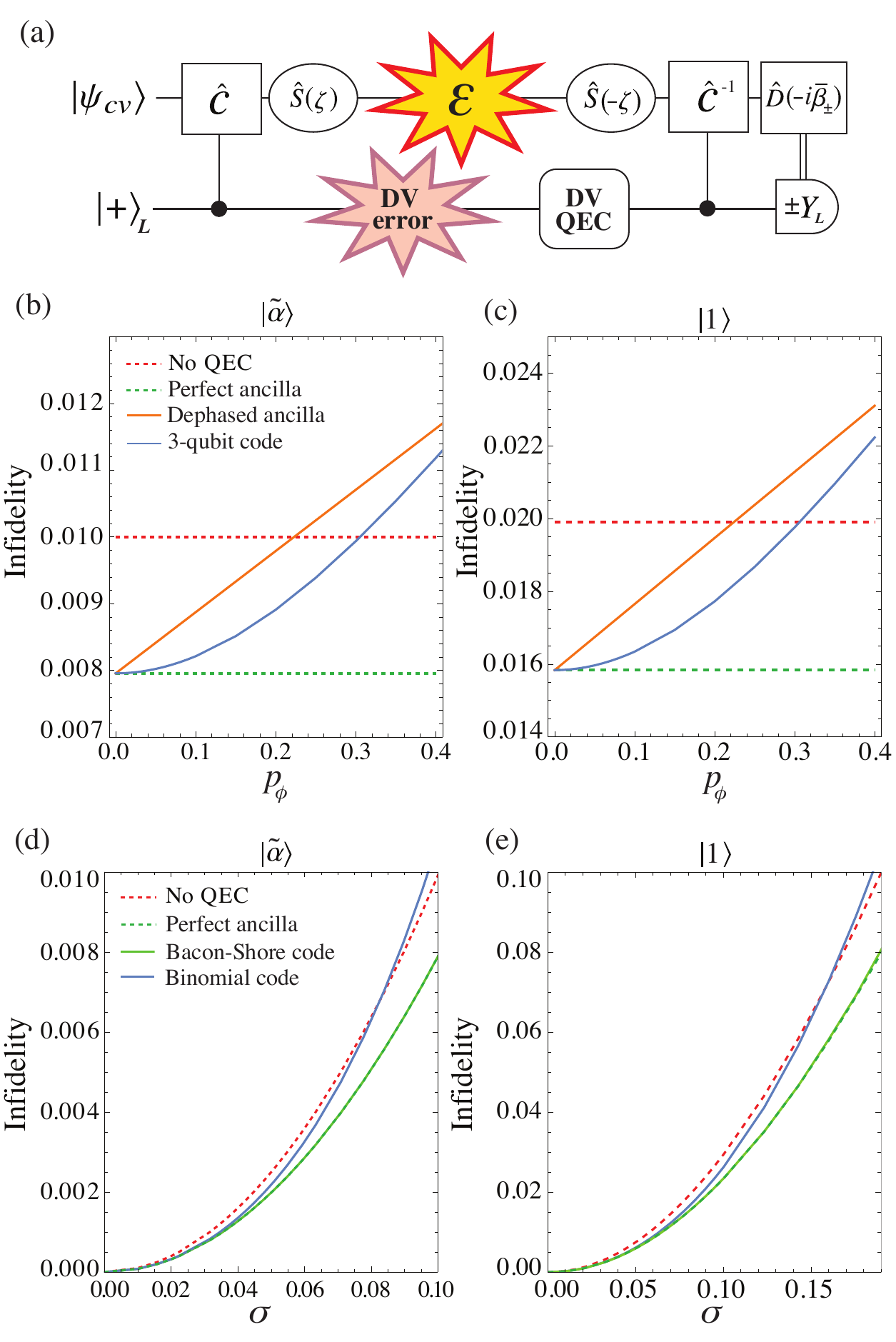}
    \caption{(a) CV QEC circuit with encoded logical qubit ancilla. 
    (b) Infidelity of Gaussian coherent state $\ket{\tilde{\alpha}}$ and (c) non-Gaussian single-boson state $\ket{1}$ when CV QEC is implemented with dephasing physical qubits.
    Here $\sigma=0.1$. 
    (d) Infidelity of $\ket{\tilde{\alpha}}$  and (e) $\ket{1}$ when CV QEC is implemented with a logical qubit in bosonic codes (\ref{eq:binomial}) and (\ref{eq:Shor}). We note that the coherent state amplitude $\tilde{\alpha}$ does not affect the infidelities in (b) and (d).
    }
    \label{fig:nonperfect}
\end{figure}

The DV QEC-concatenated CV QEC circuit is shown in Fig.~\ref{fig:nonperfect}(a). It is analogous to Fig.~\ref{fig:circuit}(a), except that the auxiliary qubit is now a logical qubit. The logical qubit will acquire error and require correction before the second conditional-displacement. As shown in Figs.~\ref{fig:nonperfect}(b) and (c), encoding the qubit ancilla can indeed improve the noise suppression for both Gaussian and non-Gaussian CV states.

\textit{Oscillator-in-oscillator code}---On the other hand, if the modes have longer coherence time than the physical qubits, such as in superconducting circuits \cite{coherence}, then we should use auxiliary modes as a bosonic-coded qudit to represent the DV ancilla \cite{10.1088/2058-9565/abe989}. The auxiliary modes should suffer from the same random displacement error (\ref{eq:noise_model}) as the mode containing CV information. 

To demonstrate the flexibility of our scheme, we consider two possible bosonic encodings. The first employs only one auxiliary mode; it represents logical qubit states by high boson states \cite{2018PhRvA..97c2346A}. Since random displacement error (\ref{eq:noise_model}) can add or remove bosons, we pick the binomial code that can correct both single-boson gain and loss \cite{2016PhRvX...6c1006M}
\begin{equation}\label{eq:binomial}
\ket{g}_L = \frac{\ket{0} + \sqrt{3}\ket{6}}{2}~,~\ket{e}_L = \frac{\sqrt{3}\ket{3}+ \ket{9}}{2}~,
\end{equation}
where $\ket{n}$ is the $n$-boson Fock state. 

Alternatively, we can use multiple modes, each utilized as a single-boson qubit, i.e. the basis states are $\ket{0}$ and $\ket{1}$. The random displacement error (\ref{eq:noise_model}) can push the auxiliary mode state outside the $\{\ket{0},\ket{1} \}$ subspace, but we can construct a dissipative process to map all higher-than-one boson states back to $\ket{1}$. Overall, a single-boson qubit state $\hat{\rho}$ is transformed as \cite{supp}
\begin{equation}
\hat{\rho} \rightarrow \frac{1}{(1+\sigma^2)^2}\hat{\rho} + \frac{\sigma^2 (2+\sigma^2)}{(1+\sigma^2)^2}\hat{\eta}~,
\end{equation}
where $\hat{\eta}\equiv \frac{1}{2+\sigma^2}\ket{0}\bra{0}+\frac{1+\sigma^2}{2+\sigma^2}\ket{1}\bra{1}$. This error process is confined in the $\{\ket{0},\ket{1} \}$ subspace, so we can use qubit codes to suppress the error. Here we use 9-qubit Shor code \cite{10.1103/physreva.52.r2493}, which can correct any single-qubit error: 
\begin{equation}\label{eq:Shor}
\ket{g}_L = \frac{(\ket{000}+\ket{111})^{\otimes 3}}{2^{3/2}}~,~\ket{e}_L = \frac{(\ket{000}-\ket{111})^{\otimes 3}}{2^{3/2}}~.
\end{equation}

The bosonic code-concatenated CV QEC also proceeds as the circuit shown in Fig.~\ref{fig:nonperfect}(a). As shown in Figs.~\ref{fig:nonperfect}(d) and (e), the CV state infidelity is suppressed in a range of $\sigma$. In particular, the Shor code outperforms the binomial code; detailed code performance analysis will be conducted in future works. To the best our knowledge, these two codes are the first known oscillator-in-oscillator codes that do not require GKP states.

\textit{Conclusion}---We have introduced a new CV QEC scheme that utilizes auxiliary DV resources. The DV ancilla extracts information about the random displacement noise, allows us to counteract the CV error. The noise suppression can be enhanced by using squeezing and a higher dimension ancilla. Our scheme can be concatenated with established DV QEC to reduce physical DV errors transferring to the CV state. It also yields a new class of oscillator-in-oscillator code that does not involve demanding GKP states. Our work provides more flexibility to implement CV QEC, thus improves the practicality of bosonic quantum technologies.


This work is supported by the Natural Sciences and Engineering Research Council of Canada (NSERC) through Discovery Grant (RGPIN-2021-02637), Alliance International Catalyst Quantum Grant (ALLRP 592649-24), CREATE (543245-2020-CREAT), Canada Research Chairs (CRC-2020-00134). E.-J. C. acknowledges hospitality of Simon Fraser University, during the stay this work was conducted.

\bibliography{reference}

\end{document}